\global\def\draftcontrol{0}

   \def\versionno{ gb-transport}

\catcode`\@=11

\expandafter\ifx\csname draftcontrol\endcsname\relax\global\def\draftcontrol{0}
\fi

{\count255=\time\divide\count255 by 60
\xdef\hourmin{\number\count255}
\multiply\count255 by-60\advance\count255 by\time
\xdef\hourmin{\hourmin:\ifnum\count255<10 0\fi\the\count255}}
\def\draftdate{\number\month/\number\day/\number\year\ \ \ \hourmin }

\newcommand\makepapertitle{\par
  \begingroup
    \renewcommand\thefootnote{\@fnsymbol\c@footnote}%
    \def\@makefnmark{\rlap{\@textsuperscript{\normalfont\@thefnmark}}}%
    \long\def\@makefntext##1{\parindent 1em\noindent
            \hb@xt@1.8em{%
                \hss\@textsuperscript{\normalfont\@thefnmark}}##1}%
     \newpage
     \global\@topnum\z@   
     \@makepapertitle
     \thispagestyle{empty}\@thanks
  \endgroup
  \setcounter{footnote}{0}%
  \global\let\thanks\relax
  \global\let\makepapertitle\relax
  \global\let\@makepapertitle\relax
  \global\let\@thanks\@empty
  \global\let\@author\@empty
  \global\let\@date\@empty
  \global\let\@title\@empty
  \global\let\title\relax
  \global\let\author\relax
  \global\let\date\relax
  \global\let\and\relax
  \def\version{\let\version\@version\@gobble}
}
\def\@makepapertitle{%
  \newpage
   \ifnum\draftcontrol=1 {}
   \version\versionno
   \vskip 3em%
   \else
   \hfill\hbox to 3cm {\parbox{4cm}{\@pubnum}\hss}%
   \vskip 3em%
   \fi
   \begin{center}%
   \let \footnote \thanks
     {\LARGE {\@title}}%
     \vskip 1.5em%
     {\normalsize
       \lineskip .5em%
       \begin{tabular}[t]{c}%
         \@author
       \end{tabular}\par}%
     \vskip 1.5em%
     {\@bstract}%
     \end{center}%
     \vskip 1.5em
     \@date%
   \par
}

\gdef\@pubnum{}
\def\pubnum#1{%
  \gdef\@pubnum{#1}}

\gdef\@bstract{}
\def\Abstract#1{%
  \gdef\@bstract{%
   \parbox{\textwidth-0pc}{%
   \centerline{\bf Abstract}\penalty1000%
\kern.2cm%
\noindent
\renewcommand\baselinestretch{1.0}%
{#1}}}
}

\def\ps@paper{\let\@mkboth\@gobbletwo%
     \ifnum\draftcontrol=1
    \def\@oddfoot{\hbox to \textwidth{\tiny \versionno \hfil\tiny\draftdate}%
    \hskip -\textwidth \hbox to \textwidth{\hfil\rm\thepage\hfil}}%
     \else\def\@oddfoot{\hbox to \textwidth{\hfil\rm\thepage\hfil}}
     \fi
     \let\@evenfoot\@oddfoot
}

\def\body{\clearpage
          \pagestyle{paper}
    }

\def\@version#1{\ifnum\draftcontrol=1
\typeout{}\typeout{#1}\typeout{}
\vskip3mm\centerline{\hbox{\fbox{\normalsize{\tt DRAFT -- #1 -- }
                   {\draftdate}}}}\vskip3mm
\fi}
\let\version\@version
\long\def\eqlabel#1{\ifnum\draftcontrol=1
                    \tag@false  
                    \tag*{(\theequation) \hbox to -0.2cm{\hspace{0cm}\small{#1}\hss}}
                    \refstepcounter{equation}
                    \edef\@currentlabel{\theequation}
                    \ltx@label{#1}          
                    \else
                    \label{#1}
                    \fi
                    }
\let\st@bibitem\@bibitem
\let\st@lbibitem\@lbibitem
\ifnum\draftcontrol=1
  \def\@bibitem#1{%
    \st@bibitem{#1}\a@@label{#1}\ignorespaces}
  \def\@lbibitem[#1]#2{%
    \st@lbibitem[#1]{#2}\a@@label{#2}\ignorespaces}
  \def\a@@label#1{%
    \gdef\a@lab{\smash{\normalfont\small#1}}
    \ifvmode
      \if@inlabel
        \global\setbox\@labels\hbox{%
          \llap{\a@lab\let\a@lab\relax
                \kern\@totalleftmargin\kern\marginparsep}%
          \box\@labels}%
      \fi
    \fi}
\fi

\documentclass[12pt,letterpaper]{article}

\usepackage{amsmath,amssymb,array,calc,epsfig,rotating,bm}
\usepackage[sort]{cite}
\usepackage{graphicx}
\usepackage{psfrag,verbatim}
\usepackage{xcolor}
\usepackage{hyperref}

\ifnum\draftcontrol=1
\fi

\renewcommand\baselinestretch{1.25}
\renewcommand\section{\@startsection {section}{1}{\z@}%
                                   {-3.5ex \@plus -1ex \@minus -.2ex}%
                                   {2.3ex \@plus.2ex}%
                                   {\normalfont\large\bfseries}}
\renewcommand\subsection{\@startsection{subsection}{2}{\z@}%
                                   {-3.25ex\@plus -1ex \@minus -.2ex}%
                                   {1.5ex \@plus .2ex}%
                                   {\normalfont\normalsize\bfseries}}
\renewcommand\subsubsection{\@startsection{subsubsection}{3}{\z@}%
                                   {-3.25ex\@plus -1ex \@minus -.2ex}%
                                   {1.5ex \@plus .2ex}%
                                   {\normalfont\normalsize\it}}
\renewcommand\paragraph{\@startsection{paragraph}{4}{\z@}%
                                   {-3.25ex\@plus -1ex \@minus -.2ex}%
                                   {1.5ex \@plus .2ex}%
                                   {\normalfont\normalsize\bf}}

\numberwithin{equation}{section}

\def\revise#1       {\raisebox{-0em}{\rule{3pt}{1em}}%
                     \marginpar{\raisebox{.5em}{\vrule width3pt\
                     \vrule width0pt height 0pt depth0.5em
                     \hbox to 0cm{\hspace{0cm}{%
                     \parbox[t]{4em}{\raggedright\footnotesize{#1}}}\hss}}}}

\newcommand\nxt[1]  {\\\fnxt#1}
\newcommand{\ie}{{\it i.e.,}\ }
\newcommand{\eg}{{\it e.g.,}\ }

\def\cala         {{\cal A}}

\def\calb         {{\cal B}}
\def\calc         {{\cal C}}
\def\cald         {{\cal D}}

\def\call         {{\cal L}}
\def\calm         {{\cal M}}

\def\calo         {{\cal O}}

\def\del          {\partial}

\def\sqr#1#2{{\vcenter{\vbox{\hrule height.#2pt
 \hbox{\vrule width.#2pt height#1pt \kern#1pt
 \vrule width.#2pt}\hrule height.#2pt}}}}

\def\dd{\delta}

\def\aa1{\phi}
\def\cc1{\psi}

\newcommand{\hq}{\mathfrak{q}}
\newcommand{\hw}{\mathfrak{w}}

\catcode`\@=12

\begin{document}


\title{\bf Holographic Gauss-Bonnet transport}

\date{February 12, 2026}

\author{
Alex Buchel\\[0.4cm]
\it Department of Physics and Astronomy\\ 
\it University of Western Ontario\\
\it London, Ontario N6A 5B7, Canada\\
\it Perimeter Institute for Theoretical Physics\\
\it Waterloo, Ontario N2J 2W9, Canada\\
}

\Abstract{
We extend the computational framework of \cite{Buchel:2023fst} to
analysis of shear and bulk viscosities in generic strongly coupled
holographic Gauss-Bonnet gauge theories.  The finite Gauss-Bonnet
coupling constant encodes holographic plasma with non-equal central
charges $c\ne a$ at the ultraviolet fixed point.  In a simple model we
discuss transport coefficients within the causality window
$-\frac12\le \frac{c-a}{c}\le \frac 12$ of the theory.
}

\makepapertitle

\body

\version\versionno
\tableofcontents

\section{Introduction and summary}\label{intro}

Until the discovery of the gauge theory/gravity correspondence \cite{Maldacena:1997re,Aharony:1999ti}
it was challenging to compute basic transport coefficients --- the shear $\eta$ and the bulk $\zeta$ viscosities ---
of gauge theory plasmas. In fact, the first reliable computations, relevant
for the quark-gluon plasma, were performed holographically  in \cite{Policastro:2001yc} for the shear viscosity,
and in \cite{Benincasa:2005iv} for the bulk viscosity. For gravitational duals  with multiple scalars,
typical in top-down holographic models of strongly coupled gauge theories, the extraction of bulk viscosity
involved computation of the dispersion relation of the sound waves\footnote{See \eg
\cite{Buchel:2005cv,Benincasa:2006ei,Buchel:2007mf,Buchel:2008uu}.}. Such computations remained technically challenging
until the Eling-Oz (EO) paper \cite{Eling:2011ms}, where the authors presented  simple formulas for the viscosities
that involved only the properties of the gravitational background, correspondingly the properties of the
holographic thermal equilibrium state. The EO formula was extensively verified in
\cite{Buchel:2011yv,Buchel:2011wx}, and shown to be equivalent to the standard holographic Kubo
method in \cite{Demircik:2023lsn}.

First holographic computations of the viscosities were performed in two-derivative gravitational duals,
correspondingly in gauge theory plasma at infinitely large 't Hooft coupling constant $\lambda=g_{YM}^2 N_c$, and for theories
with the same central charges $c=a$ at the ultraviolet fixed point. Leading-order corrections 
due to the finite 't Hooft coupling \cite{Buchel:2004di,Benincasa:2005qc,
Buchel:2008ac,Buchel:2008sh,Buchel:2008ae} and for theories with $c\ne a$
\cite{Kats:2007mq,Brigante:2008gz,Cremonini:2012ny,Buchel:2018ttd}, until recently,  
were model specific. In \cite{Buchel:2023fst} a unifying framework was presented
for the computation of transport coefficients in higher-derivative holographic models. 

We summarize now the results of \cite{Buchel:2023fst} relevant to four-derivative gravitational duals.
Consider a five-dimensional theory of gravity in AdS coupled to an arbitrary number of scalars, described by 
\begin{equation}
\begin{split}
S_5&=\frac{1}{16\pi G_N}\int_{\calm_5}d^5x \sqrt{-g}\ L_5
\\
&\equiv \frac{1}{16\pi G_N}\int_{\calm_5}d^5x \sqrt{-g}\ \biggl[
R+12-\frac{1}{2}\sum_i\left(\del \phi_i\right)^2-V\{\phi_i\}+\beta\cdot \dd\call_2
\biggr]\,,
\end{split}
\eqlabel{2der}
\end{equation}
where $\dd\call_2$ denotes the
four-derivative curvature corrections described by: 
\begin{equation}
\dd\call_2\equiv \alpha_1\ R^2+\alpha_2\ R_{\mu\nu} R^{\mu\nu}
+\alpha_3 R_{\mu\nu\rho\lambda} R^{\mu\nu\rho\lambda}\,;
\eqlabel{dl2}
\end{equation}
and the coupling constant $\beta\cdot \alpha_3$ is related to the
difference of the central charges of the UV fixed point as\footnote{There is no
simple interpretation of the remaining coupling constants $\alpha_1$ and $\alpha_2$.}
\begin{equation}
\beta\cdot \alpha_3=\frac{c-a}{8c}\,.
\eqlabel{al3ca}
\end{equation}
The shear viscosity to the entropy ratio is given by:
\begin{equation}
\frac{\eta}{s}\bigg|_{\dd\call_2}= \frac{1}{4\pi}\biggl(1+\beta\cdot
\frac23\alpha_3 \left(V-12\right)\biggr)\,;
\eqlabel{etas2}
\end{equation}
and the bulk viscosity to the entropy ratio is given by:
\begin{equation}
\begin{split}
9\pi \frac{\zeta}{s}\bigg|_{\dd\call_2}=\biggl(1&-\frac23 (V-12) (5 \alpha_1+\alpha_2-\alpha_3) \beta\biggr)
\sum_{i} z_{i,0}^2 \\&+\beta\cdot \frac{4(5 \alpha_1+\alpha_2-\alpha_3)} {3(V-12)}\ \sum_i (z_{i,0}\cdot\del_i V)^2\,,
\end{split}\eqlabel{bulks2}
\end{equation}
where $\del_i V\equiv \frac{\del V}{\del\phi_i}$.
All the quantifies in \eqref{etas2} and \eqref{bulks2}
are to be evaluated at the horizon of the
dual black brane solution. Here $z_{i,0}$ are the values of the gauge invariant
scalar fluctuations, at zero frequency, evaluated at the black brane horizon,
see appendix \ref{gbht}.

Since $\dd \call_2$ is generically higher-derivative, its coupling constant $\beta$ must be treated perturbatively.
This is an important exception though: when
\begin{equation}
\{\alpha_1\,,\, \alpha_2\,,\, \alpha_3\}=\{1\,,\, -4\,,\, 1\}\,,
\eqlabel{asb2}
\end{equation}
the combination in \eqref{dl2}
assembles into a Gauss-Bonnet term,
\begin{equation}
\dd\call_2\equiv \dd\call_{GB}\,,
\eqlabel{lgb}
\end{equation}
which renders the full gravitational action \eqref{2der}
two-derivative. In this case the coupling constant $\beta\equiv \frac{\lambda_{GB}}{2}$ can be
finite, and is constraint by the causality as \cite{Buchel:2009tt,Hofman:2008ar}    
\begin{equation}
-\frac{7}{36}\le \lambda_{GB}\le \frac{9}{100}\qquad \Longleftrightarrow\qquad -\frac 12\le \frac{c-a}{c}\le \frac 12\,.
\eqlabel{causconst}
\end{equation}
In this paper we present extensions of \eqref{etas2} and \eqref{bulks2} to holographic GB models.
We find: 
\begin{equation}
\frac{\eta}{s}\bigg|_{\dd\call_{GB}}= \frac{1}{4\pi}\biggl(1+
\frac {\lambda_{GB}}{3}\cdot  \left(V-12\right)\biggr)\,;
\eqlabel{etasgb}
\end{equation}
\begin{equation}
\begin{split}
9\pi \frac{\zeta}{s}\bigg|_{\dd\call_{GB}}=\sum_{i} z_{i,0}^2 \,,
\end{split}\eqlabel{bulkgb}
\end{equation}
valid for finite $\lambda_{GB}$. The proof of \eqref{etasgb} and \eqref{bulkgb}
is a straightforward generalization of analysis in \cite{Buchel:2023fst}.
We will omit the proof and instead refer the reader to a practical guide in
applying these formulas in appendix \ref{gbht}.

In \cite{Demircik:2024bxd} the authors studied holographic GB transport in 
models with a single bulk scalar field from entirely different perspective(s).
As we report in section \ref{sbs}, our results \eqref{etasgb} and \eqref{bulkgb},
restricted to a single scalar models, fully agree with \cite{Demircik:2024bxd}.
An intriguing claim made in \cite{Demircik:2024bxd} was that the bulk viscosity
to the entropy density obeys the unmodified EO formula:
\begin{equation}
\frac{\zeta}{s}=\frac{1}{4\pi} \left(\frac{\del\phi}{\del \ln s}\right)^2 \,,
\eqlabel{eobulk} 
\end{equation}
where the scalar field derivative with respect to the entropy density
is evaluated for its black brane horizon value.
This appears to contradict the claim made in \cite{Buchel:2023fst} that
the 'naive' EO formula is not applicable in holographic higher-derivative models.
The resolution of this is as follows. The 'naive' EO formula used in \cite{Buchel:2023fst}
was not \eqref{eobulk}, but rather
\begin{equation}
\frac{\zeta}{\eta}=\left(\frac{\del\phi}{\del \ln s}\right)^2\qquad {\rm or\ equivalently}\qquad
\frac{\zeta}{s}=\frac{\eta}{s}\cdot \left(\frac{\del\phi}{\del\ln s}\right)^2 \,.
\eqlabel{eobulkn} 
\end{equation}
For a two-derivative holographic model, \eqref{eobulk} and \eqref{eobulkn} are identical
because of the universality of the shear viscosity to the entropy density ratio in
the supergravity approximation \cite{Buchel:2003tz}
\begin{equation}
\frac{\eta}{s}=\frac{1}{4\pi}\,.
\end{equation}
The universality is lost once $\lambda_{GB}$ is nonzero \eqref{etasgb}, causing the discrepancy.
We revisit the $\dd \call_2$ models of \cite{Buchel:2023fst} in section
\eqref{eol2} and show that the bulk viscosity computed from
\eqref{bulks2} agrees with the EO formula \eqref{eobulk},
whenever the gravitational dual is effectively two-derivative
at the horizon\footnote{A holographic model was called in \cite{Buchel:2023fst}
{\it effective two-derivative at the horizon} whenever there is no difference
between the Wald and the Bekenstein entropies of the dual black brane realizing the
thermal equilibrium state of the gauge theory plasma.},
but could be higher-derivative in the bulk. If the holographic dictionary requires the use
of the Wald gravitational entropy \cite{Wald:1993nt}, the agreement between
\eqref{bulks2} and \eqref{eobulk} is lost.

\section{Applications}\label{app}

\subsection{A single bulk scalar model}\label{sbs}

As an application of holographic GB transport, consider a single scalar field model with a potential
\begin{equation}
V=\frac{m^2}{2}\phi^2\,,\qquad m^2\beta_2=\Delta (\Delta-4)\,,
\eqlabel{vap}
\end{equation}
where $\beta_2$ is given by \eqref{defl}.

\begin{figure}[ht]
\begin{center}
\psfrag{x}[c]{{$\lambda_3/T$}}
\psfrag{y}[cb]{{${\zeta}/{s}$}}
\psfrag{z}[ct]{{${4\pi\eta}/{s}$}}
  \includegraphics[width=3.0in]{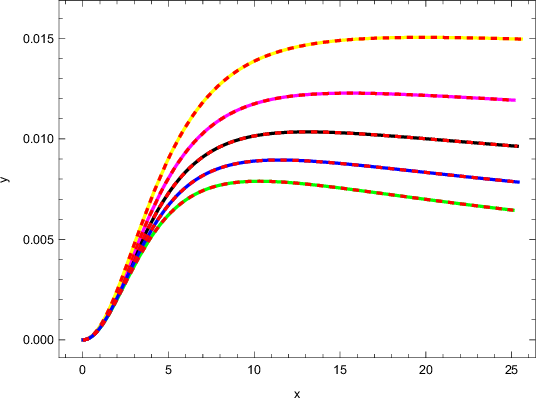}
  \includegraphics[width=3.0in]{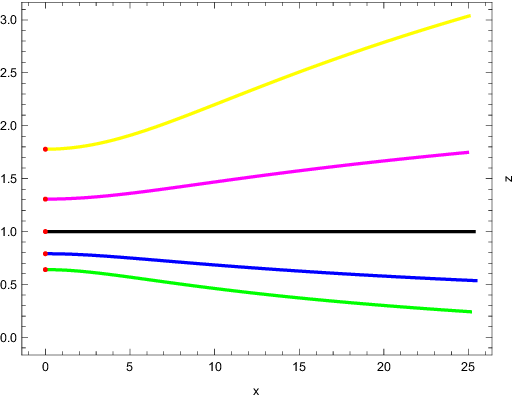}
\end{center}
\caption{Left panel: the bulk viscosity to the entropy density ratios for select values of
$(c-a)/a$, see \eqref{acsel},  in the GB holographic model \eqref{vap}. The dashed red
curves represent the corresponding ratios computed from the EO formula \eqref{eobulk}.
Right panel: the shear viscosity to the entropy density ratios in the same model.
The red dots represent the GB conformal gauge theory shear viscosity ratios \eqref{cft}.  
}\label{figure1}
\end{figure}

\begin{figure}[ht]
\begin{center}
\psfrag{x}[c]{{$\lambda_3/T$}}
\psfrag{y}[cb]{{${\zeta}/{s}$}}
\psfrag{z}[ct]{{$\zeta/\zeta^{EO}-1$}}
  \includegraphics[width=4.0in]{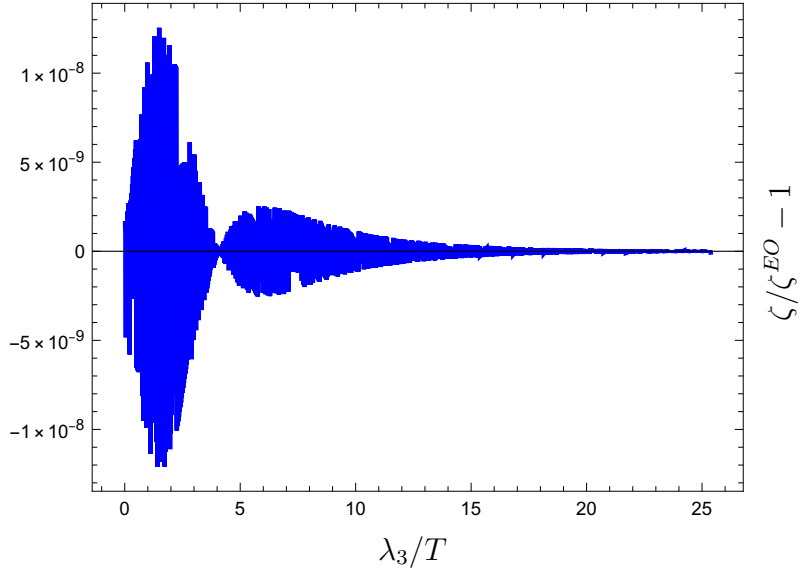}
\end{center}
\caption{There is an excellent agreement between the bulk viscosities in the GB holographic model \eqref{vap}: 
the bulk viscosity $\zeta$ is computed within the framework of
\cite{Buchel:2023fst}, and the bulk viscosity $\zeta^{EO}$ is 
extracted from the EO formula. Here we take $\frac ac=\frac 34$ (the blue curves in fig.~\ref{figure1}),
as an example.
}\label{figure2}
\end{figure}

Implementing the numerical procedure outlined in appendix \ref{gbht},
in fig.~\ref{figure1} we present the computations of the bulk (the left panel)
and the shear (the right panel) viscosities  in the single scalar model \eqref{vap}
for select values of $(c-a)$ within the causality range \eqref{causconst}.
The color coding of the solid curves is as in
\begin{equation}
\frac{c-a}{c}\ =\ \biggl\{{\color{yellow}{-\frac12}}\,,\, {\color{magenta}-\frac 14}\,,\,
{\color{black} 0}\,,\, {\color{blue} \frac 14}\,,\ {\color{green}\frac 12}\biggr\}\,.
\eqlabel{acsel}
\end{equation}
The dashed red lines are the computations of the bulk viscosity using the EO formula \eqref{eobulk}.
The red dots represent the holographic GB conformal shear viscosity to the entropy density ratios
\cite{Kats:2007mq,Brigante:2008gz}:
\begin{equation}
\frac{4\pi \eta}{s}\bigg|_{\lambda_3=0}=1-4\lambda_{GB}=\frac{4}{\left(\frac ac-3\right)^2}\,.
\eqlabel{cft}
\end{equation}
Note that $\frac{4\pi\eta}{s}$ ratio (the green curve)
is always below the minimal conformal
value $\frac{16}{25}$ \cite{Brigante:2008gz},
even without the phase transition that
decouples the microcausality (the UV property) of the theory from
its hydrodynamic transport (the IR property) \cite{Buchel:2010wf}.

There is an excellent numerical agreement between the bulk viscosity
results from \eqref{bulkgb} and \eqref{eobulk}:
in fig.~\ref{figure2} we show a typical difference as exemplified by $(c-a)/c=\color{blue} 1/4$
(the blue curve in the left panel of fig.~\ref{figure1}).

\subsection{EO formula in $\dd \call_2$ models}\label{eol2}

In this section we revisit $\dd \call_2$ models of \cite{Buchel:2023fst} with respect to comparison
with an {\it inequivalent } EO formula \eqref{eobulk}. We focus
on\footnote{Our conclusions extend to all the other models in \cite{Buchel:2023fst}. } $\cala_{2,3}$,
$\calb_{2,3}$ and $\calc_{2,3}$ models discussed there. In all these models the bulk scalar field is
dual to a dimension $\Delta=3$ boundary gauge theory operator.
The sets of coupling constants $\alpha_i$ in \eqref{2der} are as follows:
\begin{itemize}
\item model $\cala$: $\qquad \{\alpha_1\,,\, \alpha_2\,,\, \alpha_3\}=\{0\,,\, 0\,,\, 1\}$; 
\item model $\calb$: $\qquad \{\alpha_1\,,\, \alpha_2\,,\, \alpha_3\}=\{1\,,\, -4\,,\, 1\}$; 
\item model $\calc$: $\qquad \{\alpha_1\,,\, \alpha_2\,,\, \alpha_3\}=\{0\,,\, 1\,,\, 1\}$. 
\end{itemize}
Model $\calb$ is just the model of section \ref{sbs} to the leading order
in the GB coupling constant; model $\calc$ has a peculiar property that
while it is higher-derivative in the bulk, the corresponding black
brane entropy density is given by the Bekenstein formula ---
it is effectively two-derivative in the vicinity of the horizon.
Finally, model $\cala$ is genuinely higher-derivative:
the correct thermodynamic entropy of the boundary
gauge theory thermal state must be identified with the Wald entropy
of the dual black brane.

\begin{figure}[ht]
\begin{center}
\psfrag{x}[c]{{$m_f/(2\pi T)$}}
\psfrag{t}[cb]{{$\dd_{\zeta,3}^\calb$}}
\psfrag{z}[ct]{{$\dd_{\zeta,3}^\calc$}}
  \includegraphics[width=3.0in]{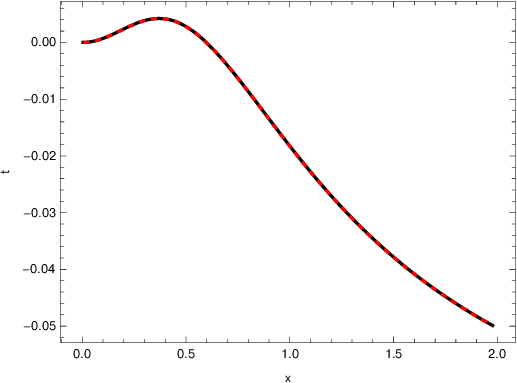}
  \includegraphics[width=3.0in]{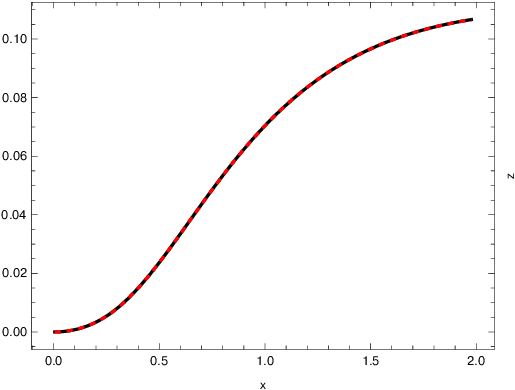}
\end{center}
\caption{At order  $\calo(\beta)$ there is a  perfect agreement
between the ratio of the bulk viscosity to the
entropy density evaluated using  \eqref{bulks2} (the solid black curves)
and the extension of  the EO formula \eqref{eobulk}  to order $\calo(\beta)$
(shown in the red dashed curves), in
holographic models which are effectively two-derivative at the horizon. 
}\label{figure3}
\end{figure}

As shown in fig.~\ref{figure3},  there is now an agreement
for models $\calb$ and $\calc$ for the
two methods of computing the bulk viscosity to the
entropy density ratio: using \eqref{bulks2} (which actually
reduces to \eqref{etasgb} for both models) and the EO formula 
\eqref{eobulk}. Of course, the agreement for model $\calb$
is expected. The agreement for model $\calc$
(and similar models of \cite{Buchel:2023fst}) strongly suggests
that the EO formula \eqref{eobulk} is more widely applicable: it is
correct in higher-derivative holographic models that do not distinguish
between the Wald and the Bekenstein entropies of the dual black brane.

\begin{figure}[ht]
\begin{center}
\psfrag{x}[c]{{$m_f/(2\pi T)$}}
\psfrag{t}[cb]{{$\dd_{\zeta,3}^\cala$}}
\psfrag{z}[ct]{{$\dd c_s^2$}}
  \includegraphics[width=3.0in]{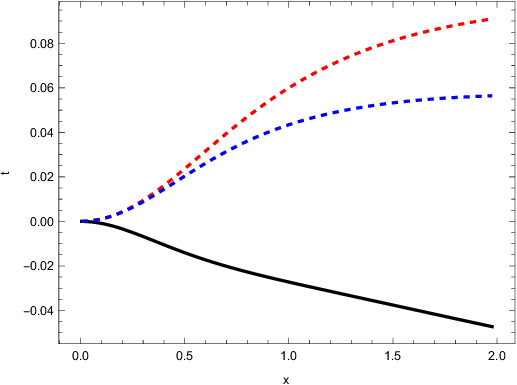}
  \includegraphics[width=3.0in]{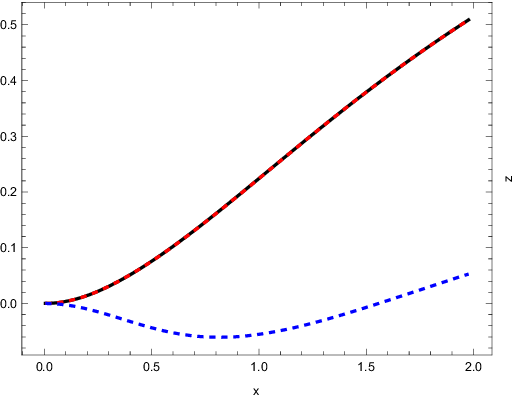}
\end{center}
\caption{Left panel: at order $\calo(\beta)$ there is a disagreement 
between the ratio of the bulk viscosity to the
entropy density evaluated using  \eqref{bulks2}
(the solid black curve, which is independently verified in \cite{Buchel:2023fst},
using the sound wave attenuation computation),
and the extensions of  the EO formula \eqref{eobulk}  to order $\calo(\beta)$,
where the boundary gauge theory
thermal state entropy density $s$ is identified with the
dual black brane Wald entropy $s=s_{Wald}$ (the red dashed curve),
or when the boundary gauge theory
thermal state entropy density is identified with the
dual black brane Bekenstein entropy $s=s_{Bekenstein}$ (the blue dashed curve).
Right panel: $\calo(\beta)$ correction to the speed of the
sound waves extracted from the sound wave dispersion relation \eqref{sdisp}
is represented by solid black curve; the computation
of the same quantity from the thermodynamic relation \eqref{csth}
with $s=s_{Wald}$ is shown in the red dashed curve, and with $s=s_{Bekenstein}$
is shown in the blue dashed curve. This validates the holographic
dictionary $s\equiv s_{Wald}$.
}\label{figure4}
\end{figure}

The agreement between \eqref{bulks2} and \eqref{eobulk} is lost
for model $\cala$, see fig.~\ref{figure4}. It is so when one either
uses the dual black brane Bekenstein entropy density $s=s_{Bekenstein}$
in \eqref{eobulk} (the dashed blue curve), or its Wald
entropy density $s=s_{Wald}$ (the dashed red curve).
Of course, the holographic dictionary requires to identify
$s=s_{Wald}$. This can be independently verified in model $\cala$
by extracting the speed of the sound waves $c_s^2$ from the
dispersion relation (this computation was performed in \cite{Buchel:2023fst})
\begin{equation}
\hw=c_s\cdot \hq  +\calo(\hq^2)\,,\qquad c_s\equiv \frac{\beta_{1,0}}{\sqrt{3}}
\biggl(1+\beta\cdot \beta_{1,1}\biggr)\,,\qquad
\dd c_s^2\equiv \frac{2 \beta_{1,0}^2 \beta_{1,1}}{3}\,,
\eqlabel{sdisp}
\end{equation}
and comparing the result with the sound speed extracted
from the thermodynamics:
\begin{equation}
c_s^2=\frac{\del\ln T}{\del\ln s}\,.
\eqlabel{csth}
\end{equation}
In the right panel of fig.~\ref{figure4} we show $\dd c_s^2$
from \eqref{sdisp} (the solid black curve)
and the same quantity extracted from \eqref{csth}
with the identification $s=s_{Bekenstein}$ (the blue dashed curve) and
with the identification $s=s_{Wald}$ (the red dashed curve).

\section*{Acknowledgments}
I would like to thank Sera Cremonini and Umut Gursoy for
valuable discussions.
Research at Perimeter
Institute is supported by the Government of Canada through Industry
Canada and by the Province of Ontario through the Ministry of
Research \& Innovation. This work was further supported by
NSERC through the Discovery Grants program.

\appendix

\section{Practical guide to the GB holographic transport}\label{gbht}

\subsection{The background equations of motion}

The background geometry dual to a thermal equilibrium state
of a boundary gauge theory is 
\begin{equation}
ds_5^2=-c_1^2\ dt^2+c_2^2\ d\bm{x}^2+ c_3^2\ dr^2\,,
\eqlabel{5metric}
\end{equation}
where $c_i=c_i(r)$, and additionally $\phi_i=\phi_i(r)$.
The radial coordinate is $r\in [0,r_h]$, with $r_h$ being the
location of the regular black brane horizon,
\begin{equation}
\lim_{r\to r_h} c_1=0\,.
\eqlabel{defrh}
\end{equation}

From \eqref{2der} we obtain the following equations of
motion:
\begin{equation}
\begin{split}
&0=c_1''-\frac{c_1' c_3'}{c_3}+\frac{c_2^2 c_3^2}{12 \cald^2} \biggl(
c_1 c_2^2 c_3^2-4 c_2' \beta (3 c_1 c_2'-2 c_2 c_1')
\biggr) \biggl(2 c_3^2 (V-12)+\sum_j(\phi_j')^2\biggr)
\\&-\frac{c_2' c_3^2}{\cald^2} \biggl(
c_1 c_2^2 c_3^2 c_2'-2 c_2^3 c_3^2 c_1'+4 c_1 (c_2')^3 \beta
\biggr)\,,
\end{split}
\eqlabel{eq1}
\end{equation}
\begin{equation}
\begin{split}
&0=c_2''+\frac{c_2^3 c_3^2}{12 \cald} \biggl(
2 c_3^2 (V-12)+\sum_j (\phi_j')^2\biggr)
-\frac{c_2'}{c_3 \cald} \biggl(
c_2^2 c_3^2 c_3'-c_2 c_3^3 c_2'-4 (c_2')^2 c_3' \beta\biggr)\,,
\end{split}
\eqlabel{eq2}
\end{equation}
\begin{equation}
\begin{split}
&0=\sum_i(\phi_i')^2 -\frac{12 (c_2')^2}{c_2^2}-\frac{12 c_2' c_1'}{c_2 c_1}
-2 c_3^2 (V-12)+\frac{48 c_1' (c_2')^3 \beta}{c_1 c_2^3 c_3^2}\,,
\end{split}
\eqlabel{eq3}
\end{equation}
\begin{equation}
\begin{split}
&0=\phi_i''-\frac{\phi_i' c_3'}{c_3}+\frac{3 \phi_i' c_2'}{c_2}+\frac{c_1' \phi_i'}{c_1}-c_3^2\cdot \del_i V\,,
\end{split}
\eqlabel{eq4}
\end{equation}
where we defined
\begin{equation}
\cald=c_2^2 c_3^2-4 (c_2')^2 \beta\,.
\eqlabel{dd}
\end{equation}
We verified that the constraint \eqref{eq3} is consistent with the remaining equations at finite $\beta$.

On shell, \ie evaluated when \eqref{eq1}-\eqref{eq4} hold, the effective action
\eqref{2der}
is a total derivative. Specifically,
we find 
\begin{equation}
\sqrt{-g}L_5=-\frac{6c_1\cald}{c_3^3}\ \cdot\ {\rm eq.}\eqref{eq2}
+\frac{d}{dr}\biggl\{\frac{-2 c_1'c_2}{c_3^3}
\biggl(c_2^2 c_3^2-12 (c_2')^2 \beta\biggr)\biggr\}\,.
\eqlabel{totder}
\end{equation}
Following \cite{Buchel:2023fst}, we identify
\begin{equation}
s T = \frac{1}{16\pi G_N}\lim_{r\to r_h}\biggl[\frac{-2 c_1'c_2}{c_3^3}
\biggl(c_2^2 c_3^2-12 (c_2')^2 \beta\biggr)
\biggr]\,,
\eqlabel{defst}
\end{equation}
where $s$ and $T$ are the entropy density and the temperature of the
boundary gauge theory thermal state.

For numerical analysis it is convenient to parameterize the GB coupling constant $\lambda_{GB}$ as
\begin{equation}
\lambda_{GB}\equiv 2\beta=\beta_2-\beta_2^2\,.
\eqlabel{defl}
\end{equation}
It is also useful to adopt the radial gauge as
\begin{equation}
c_1\equiv \frac{\sqrt{\beta_2 f_1 f_2}}{r}\,,\qquad c_2=\frac 1r\,,\qquad c_3=\frac{1}{r\sqrt{f_1}}\,,
\eqlabel{ci}
\end{equation}
where $r\in (0,r_h\equiv 1]$.
The background equations of motion \eqref{eq1}-\eqref{eq4} are solved subject to the following
asymptotics:
\nxt in the UV, \ie as $r\to 0_+$,
\begin{equation}
f_1=\frac 1\beta_2+\cdots\,,\qquad f_2=1+\cdots\,,\qquad \phi_i=\lambda_i\ r^{4-\Delta_i}+\cdots\,,
\eqlabel{uvb}
\end{equation}
where $\cdots$ indicates the subleading terms near the asymptotically $AdS_5$ boundary,
and $\lambda_i$ are the non-normalizable coefficients for the bulk
scalar fields $\phi_i$ dual to the couplings of relevant and marginal boundary gauge theory operators $\calo_i$
of conformal dimension $\Delta_i$, 
\begin{equation}
m_i^2 \beta_2=\Delta_i (\Delta_i-4)\,,
\eqlabel{mass}
\end{equation}
with $m_i$ being the bulk mass\footnote{See \cite{Buchel:2018ttd} for the
holographic dictionary $m_i^2\leftrightarrow \Delta_i$ when $\beta\ne 0$.} of $\phi_i$;
\nxt in the IR, \ie as $r\to r_h=1$,
\begin{equation}
f_1=0+\cdots\,,\qquad f_2={\rm finite}+\cdots\,,\qquad \phi_i={\rm finite}+\cdots\,.
\eqlabel{irb}
\end{equation}

In the parameterization \eqref{ci} one obtains 2 first-order equations for $f_1$ and $f_2$
(from \eqref{eq2} and \eqref{eq3}) and $n$ second-order equations for the bulk
scalars $\phi_i$. For a fixed set of the non-normalizable coefficients $\{\lambda_i\}$,
there are in total $2\times n+2$ parameters: the normalizable coefficient of $f_1$ (related to the
energy density of the boundary gauge theory plasma), the horizon value of $f_2$,
the $n$ normalizable coefficients of $\phi_i$, and the $n$ horizon values of $\phi_i$. 
It is important to keep in mind that fixing $c_2$ as in \eqref{ci} sets the overall entropy scale;
thus it is important to parameterize/extract all the data as dimensionless quantities, \eg
\begin{equation}
\frac{\lambda_i}{T^{4-\Delta_i}}\,,\qquad \frac{s}{T^3}\,,
\eqlabel{lpar}
\end{equation}
where $T$ is the Hawking temperature,
\begin{equation}
T=\lim_{r\to r_h}\ \biggl\{-\frac{c_2}{2\pi c_3}\left(\frac{c_1}{c_2}\right)'\biggr\}\,,
\eqlabel{deft}
\end{equation}
and $s$ is the entropy density\footnote{In GB holography the black brane
Wald entropy is the same as its Bekenstein entropy. It also follows from
\eqref{defst}, given \eqref{deft}.},
\begin{equation}
s=\lim_{r\to r_h}\frac{c_2^3}{4 G_N}=\frac{1}{4 G_N} \,.
\eqlabel{defs}
\end{equation}

The shear viscosity to the entropy density in the model \eqref{2der} can be extracted
once the background solution is constructed --- from \eqref{etasgb} it depends only on the
value of the GB coupling constant $\lambda_{GB}=2\beta$, and the value of the bulk scalar potential
evaluated at the black brane horizon, \ie on the value of
\begin{equation}
\lim_{r\to r_h} V\{\phi_i\} = V\left\{\lim_{r\to r_h} \phi_i\right\}\,.
\eqlabel{vh}
\end{equation}

\subsection{Equations of motion for $z_{i,0}$}

As detailed in \cite{Buchel:2023fst}, $z_{i,0}$ are the gauge-invariant scalar fluctuations
at zero spatial momentum and zero frequency.
They represent the decoupled set of the sound channel quasinormal modes. 
For the model \eqref{2der} we find
\begin{equation}
\begin{split}
&0=z_{i,0}''+\left(\ln\frac{c_1 c_2^3}{c_3}\right)' z_{i,0}'
+\sum_j z_{j,0}\cdot\biggl\{
-\frac{\phi_i' c_2^3 c_3^4}{3c_2' \cald}\cdot  \del_jV-c_3^2 \del^2_{ij}V
\\&-\frac{c_3^4 c_2^6 \phi_i' \phi_j'}{18 (c_2')^2 \cald^3} \biggl(c_2^2 c_3^2-8 (c_2')^2 \beta\biggr)\cdot
\sum_k (\phi_k')^2\\&+\phi_j'\cdot \biggl[
-\frac{c_2^3 c_3^4}{3\cald c_2'}\cdot \del_iV+\frac{2 \phi_i' c_2^3 c_3^2}{3\cald^2 c_1 c_2'}
\biggl(
c_1 c_2 c_3^2 c_2'+c_2^2 c_3^2 c_1'-8 (c_2')^2 c_1') \beta\biggr)
\biggr]
\biggr\}\,,
\end{split}
\eqlabel{zi0}
\end{equation}
where $\cald$ is given by \eqref{dd}.

The set of equations \eqref{zi0} is solved subject to the following asymptotics:
\nxt in the UV\footnote{See \cite{Buchel:2023fst} for detailed discussion
of the UV boundary conditions.}, \ie as $r\to 0_+$,
\begin{equation}
z_{i,0}=\frac{4-\Delta_i}{2}\cdot \lambda_i\ r^{4-\Delta_i}+\cdots\,,
\eqlabel{zuv}
\end{equation}
where $\cdots$ indicates the subleading terms near the asymptotically $AdS_5$ boundary;
\nxt in the IR, \ie as $r\to r_h=1$,
\begin{equation}
z_{i,0}=\underbrace{z_{i,0}^h}_{\rm finite}+\cdots\,.
\eqlabel{zir}
\end{equation}

There are in total $n$ (the same number as the number of bulk scalar fields)
second-order equations in the set \eqref{zi0} --- the solution is
fully specified by the $n$ normalizable coefficients of the fluctuations $z_{i,0}$, and their $n$
horizon values $z_{i,0}^h$. From \eqref{bulkgb}, the ratio of the bulk viscosity to the entropy density
is given by
\begin{equation}
9\pi \frac{\zeta}{s}=\sum_{i} (z_{i,0}^h)^2\,.
\eqlabel{finf}
\end{equation}

\bibliographystyle{JHEP}
\bibliography{gbt}

\end{document}